\documentclass[11pt,twoside]{article}
\usepackage{asp2010}

\resetcounters

\begin{document}

\title{Practices in Code Discoverability: Astrophysics Source Code Library}
\author{Alice Allen$^1$, Peter Teuben$^2$, 
Robert J. Nemiroff$^3$, and Lior Shamir$^4$
\affil{$^1$Calverton, MD}
\affil{$^2$Astronomy Department, University of Maryland}
\affil{$^3$Michigan Technological University}
\affil{$^4$Lawrence Technological University}}

\begin{abstract}

Here we describe the Astrophysics Source Code Library (ASCL), which
takes an active approach to sharing astrophysical source code. ASCL's
editor seeks out both new and old peer-reviewed papers that describe
methods or experiments that involve the development or use of source
code, and adds entries for the found codes to the library. This
approach ensures that source codes are added without requiring authors
to actively submit them, resulting in a comprehensive listing that
covers a significant number of the astrophysics source codes used in
peer-reviewed studies. The ASCL now has over 340 codes in it and
continues to grow.

In 2011, the 
ASCL\footnote{\url{http://asterisk.apod.com/viewforum.php?f=35}}
has on
average added 19 new codes per month. An advisory committee has been
established to provide input and guide the development and expansion
of the new site, and a marketing plan has been developed and is being
executed. All ASCL source codes have been used to generate results
published in or submitted to a refereed journal and are freely
available either via a download site or from an identified source.

This paper provides the history and description of the ASCL. 
It lists the requirements for including codes, examines the benefits 
of the ASCL, and outlines some of its future plans.

\end{abstract}

Note: we don't seem to have a reference \citep{O27_adassxxi} for our other paper.

\section{History of the ASCL}

In 1999, Robert Nemiroff and John F. Wallin founded the
online Astrophysics Source Code Library (ASCL) to house codes
of use to the community.\cite{1999AASÉ194.4408N} This
was a volunteer, spare-time endeavor and resulted in a library of 37
codes which had been described in the literature and used to produce
research published in or submitted to refereed journals. Twenty-seven
of the codes were added to the library in 1999; the last code was
added in 2002. The ASCL site also linked to other code libraries, most
of which no longer exist or have not been updated in years. In 2003, 
a search for a new editor for the ASCL was unsuccessful. As other code 
libraries appeared to be under development at that time, the ASCL 
was no longer being updated though remained available.

In 2010, Nemiroff decided to move the information on the old ASCL site to 
Starship Asterisk, the discussion forum for APOD.
\footnote{Starship Asterisk:
\url{http://asterisk.apod.com/viewtopic.php?f=29&t=23735}} He enlisted
volunteer help to move entries to the new site and expand the library
(Nemiroff, 2010). In 2011, an advisory committee was established 
to provide guidance for the development and expansion of the new site.

\section{Description of the ASCL and code entry requirements}

Starship Asterisk runs on the widely-used open
source bulletin board software phpBB. Its index page has two main sections, one of
which is {\it Learning \& Resources.} The ASCL is housed in this
section in a separate forum called {\it The Engineering
Deck: Astrophysics Source Code Library}. 

The first three threads of the ASCL are informational threads rather than code 
threads and include how to add a code, which codes have recently been added, 
and papers and other resources which may be of interest to astrophysicists 
and astronomers. Each code listed in the ASCL has its own thread; 
the first post of
a code thread contains the following information:

\begin{itemize}
\item Name of code
\item Abstract or description of code
\item Person(s) credited with writing the code
\item Link to the source code site
\item Link to a paper which discusses or uses the code
\item Unique number for the code
\end{itemize}

Figure~\ref{STIFF} shows a code entry, annotated for the ADASS XXI poster presentation on the ASCL.

\articlefigure{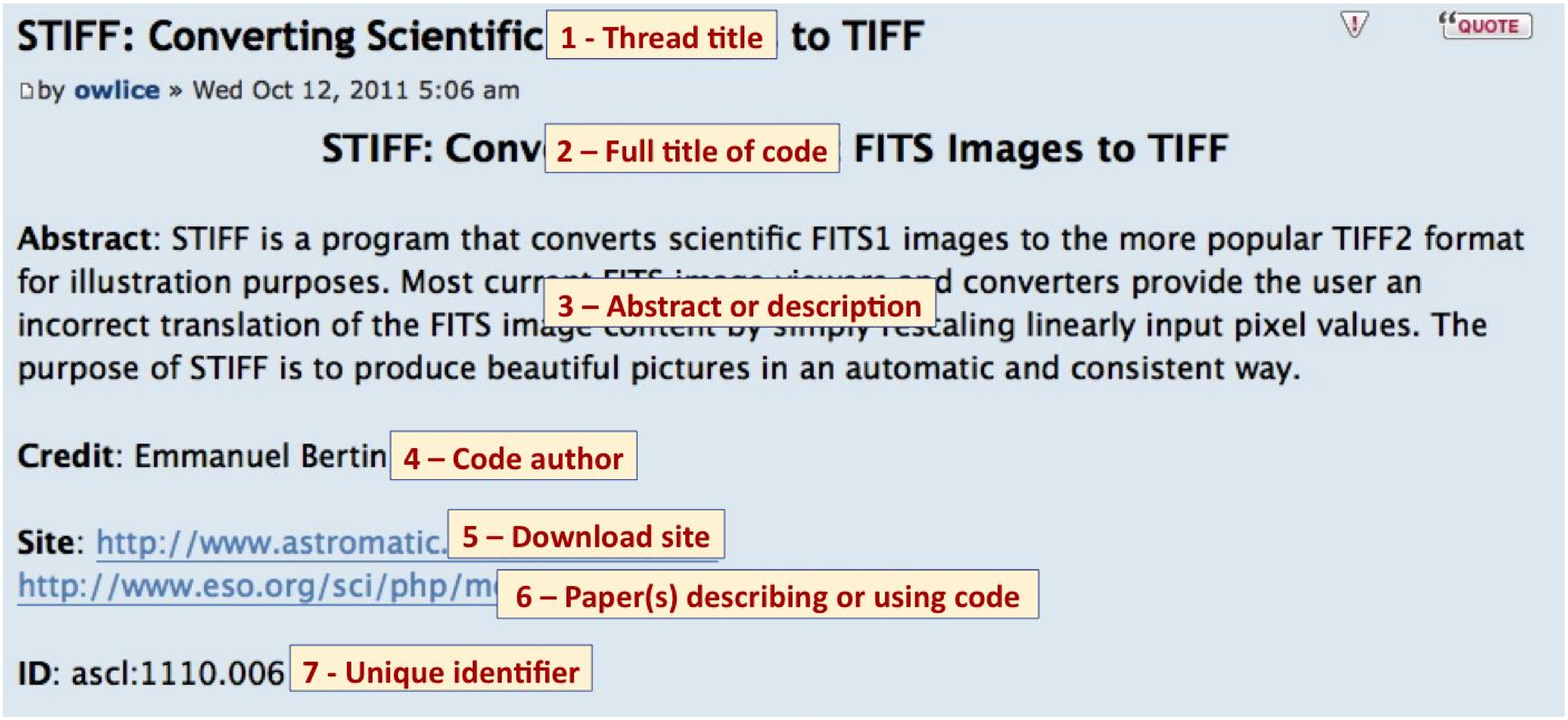}{STIFF}
{Annotated code entry; this graphic is taken from the poster presented at ADASS XXI. The abstract for the code was condensed for display.}

Questions about and discussion of the code can be posted to the thread
by clicking the POST REPLY button (not shown in Fig.~\ref{STIFF}) at the top or bottom of the
post. It is not necessary to register for the Starship Asterisk forum
to read and post on the ASCL site, however, there is an advantage to
doing so: registered users can subscribe to the ASCL forum and/or a
particular thread on the forum; subscribing alerts a user via email
when the thread or forum has been updated. 

Though most entries currently do not house the codes themselves, it is
possible to attach an archive file (i.e., .zip or .gz) to the code
entry for downloading.

Codes are listed alphabetically by name, 100 threads
to a page. A full-text search capability is available, and searches can be
refined by iterative searching on the results. The ability to search
will become increasingly important as the library grows.

\section {Benefits of the ASCL}

Each of the various efforts to aid communication and share knowledge of codes
useful for astrophysics has offered valuable information to the
community; the difficulty lies in informing and reminding the
community that the resource exists. According to Nemiroff, the
original ASCL was not successful because {\it most people just didn't
know about it and had few ways to find it.}

He came to realize that consistent exposure is needed for a resource
to become known and used. With APOD as an
entry point to the ASCL, APOD can create the exposure needed to inform
and remind astrophysicists of this resource. This is done by posting a
link periodically from APOD to the ASCL.

Additional notice about the ASCL is provided by an ongoing
email campaign to inform coders of the ASCL and requests to sites
which link to the old ASCL site to link to the new one though the old site 
redirects to the new.

Because the editor seeks codes from peer-reviewed papers and adds entries 
for them to the library, the ASCL currently houses the largest collection 
of codes known to the authors. The platform on which the ASCL is housed is 
familiar and easy to use; it allows for discussion of each code on its own 
thread, attachment of archive files for those codes which do not have 
download sites of their own, and consistent updating and expansion. 
The alphabetical listing of the code entries and full-text and iterative 
search capabilities allow users to find codes of interest quickly.

\section {Growth, usage, and future plans}

The ASCL has been expanded greatly from last year to this, and as of
this writing, has 340 codes in it. The ASCL has seen 808\% growth 
in the number of codes from the 3rd quarter of 2010 to the 3rd quarter 
of 2011, as Figure~\ref{Graphs} shows. The expansion will continue; we
currently have nearly 200 codes in the queue to be added, and actively
seek newly-released codes to include. We invite the astrophysics
community to suggest codes that are missing; codes that the community
requests be added move to the front of the queue for inclusion.

As the resource has received exposure through APOD, the email campaign, and 
posts on blogs such as AstroBetter\footnote{AstroBetter:
\url{http://www.astrobetter.com/}} and Astronomy Computing Today,\footnote{Astronomy Computing Today:
\url{http://astrocompute.wordpress.com/}} visits to 
the ASCL have increased; this is demonstrated by the bar graph, of two 30-day periods in 2011, in Figure~\ref{Graphs}. 

\articlefigure{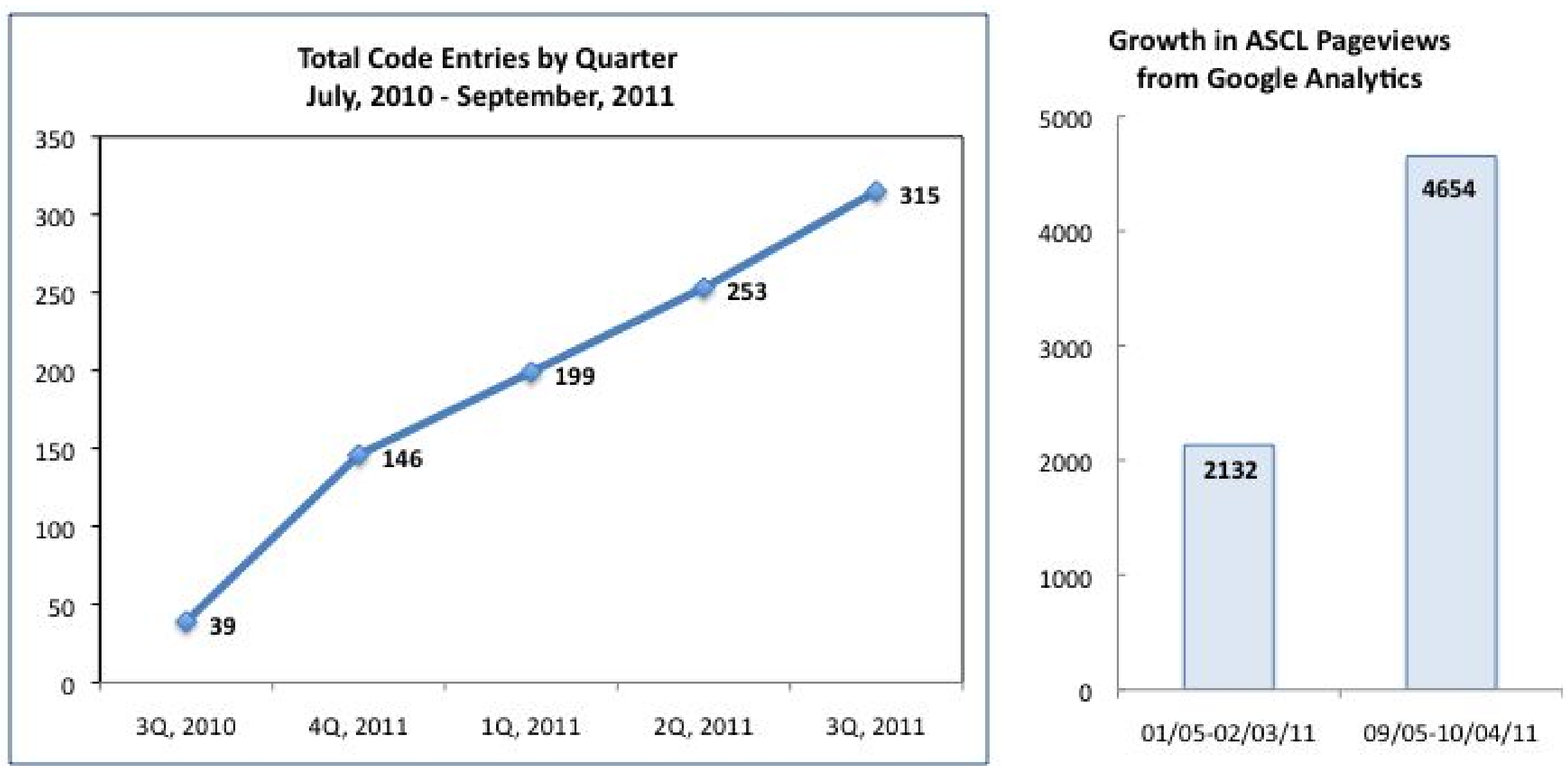}{Graphs}{Left: From the 3rd quarter 
of 2010 through the 3rd quarter of 2011, the ASCL grew from 39 to 315 entries.
As of this writing (October, 2011), the ASCL has 340 entries in it.
Right: Visits to the ASCL have increased from the beginning of 2011; here, 
two 30-day periods show a 118\% increase in the number of visits to the ASCL, 
as determined by Google Analytics}

We are
exploring ways to make the ASCL citable; we believe papers which
use codes should cite them, and are working to provide an easy method
for doing so.

\section {Conclusion}

Because of the depth and breadth of the ASCL, the
ongoing work to expand it, the exposure provided through APOD, the
guidance of advisory committee members who know the astronomical
coding community well, and the ease of using the phpbb platform, we feel the
newly revised ASCL will become a valuable resource for astronomers and
astrophysicists.

{\bf NOTE ADDED IN PROOF:}  ASCL codes are now 
incoorporated into ADS.

\bibliography{P003}

\end{document}